\documentclass[letterpaper,10pt]{article} 

\usepackage{osameet3} 

\usepackage{amsmath,amssymb}
\usepackage{upgreek}
\usepackage{multirow}
\usepackage{pgfplots}
\usepackage{graphicx}
\usepackage{subcaption}
\usepackage{mwe}
\usepackage{adjustbox}
\usepackage{xcolor}
\usepackage{collcell}
\usepackage{multirow}
\usepackage{mathtools}
\usepackage[colorlinks=true,bookmarks=false,citecolor=blue,urlcolor=blue]{hyperref} 

\begin{document}

\title{Experimental Study of Deep Neural Network Equalizers Performance in Optical Links}

\author{Pedro J. Freire\textsuperscript{(1)}, Yevhenii Osadchuk\textsuperscript{(2)}, Bernhard Spinnler\textsuperscript{(3)}, Wolfgang Schairer\textsuperscript{(3)}, Antonio Napoli\textsuperscript{(3)}, Nelson Costa\textsuperscript{(4)}, Jaroslaw E. Prilepsky \textsuperscript{(1)}, Sergei K. Turitsyn\textsuperscript{(1)}}
\address{\textsuperscript{(1)}~Aston University, Birmingham, UK \textsuperscript{(2)} National and Kapodistrian University of Athens, Greece
   \textsuperscript{(3)}~Infinera, Munich, Germany  \textsuperscript{(4)}  Infinera Unipessoal, Carnaxide, Portugal
  
   }
  \vspace{-0.5mm}
\email{p.freiredecarvalhosourza@aston.ac.uk}

\copyrightyear{2021}


\begin{abstract}
We propose a convolutional-recurrent channel equalizer and experimentally demonstrate 1dB Q-factor improvement both in single-channel and 96$\times$WDM, DP-16QAM transmission over 450km of TWC fiber. The new equalizer outperforms previous NN-based approaches and a 3-steps-per-span DBP.
\end{abstract}



\section{Introduction}
\vspace{-1.5mm}

The application of machine learning (ML) and neural networks (NN) for signal's distortion compensation in optical links has become a subject of intensive study, largely due to the NNs' capability of reverting the nonlinear channel with high noise tolerance~\cite{hager2020physics,zhang2019field,sidelnikov2019methods,deligiannidis2020compensation}. In particular, numerical modeling of a multi-layer perceptron (MLP)~\cite{sidelnikov2019methods} in a long-haul scenario demonstrated a bit error rate (BER) improvement similar to the one rendered by digital back-propagation (DBP) with 2 steps-per-span (StPS) and 2 samples-per-symbol. More advanced NN architectures, such as bidirectional Long Short-Term Memory (biLSTM) NNs~\cite{deligiannidis2020compensation} were also simulated over metro and long-haul links demonstrating better performance than the one obtained by DBP with 6 StPS.

In this study, we examine the performance of the previously proposed NN equalizers, the MLP, and biLSTM, using experimental data. We introduce a novel scheme of the equalizer based on the convolutional-recurrent NN (CRNN) technique that combines the properties of convolutional and recurrent layers. The performance of the proposed equalizer is tested and compared to other NN architectures and DBP~\cite{dbp2014} in an experimental setup. We analyze the single-channel (SC) and wavelength-division multiplexing (WDM) transmission of a dual-polarization (DP) 16 QAM signal with 34.4~GBd rate along 9$\times$50km TrueWave Classic (TWC) fiber spans. First, the CRNN is evaluated in an SC case providing about 1~dB improvement in Q-factor when compared to a traditional DSP~\cite{kuschnerov2010data}. Next, we investigate the performance of the new equalizer in the WDM scenario with 96 channels. In this experiment, we specifically aim to identify whether the biLSTM is capable to compensate for any of the cross-phase modulation (XPM) distortions, as it was observed numerically in \cite{deligiannidis2020compensation}. Our results confirm that the NN equalizer architectures that involve a biLSTM layer are able to partially compensate for the XPM induced distortions. In the WDM case, the proposed CRNN equalizer increases the maximum Q-factor by up to 1.04~dB, with the optimal launch power increasing by 2 dB. In both scenarios, the CRNN equalizer outperforms the results delivered by the MLP and biLSTM, also showing better performance than the traditional DBP with 3 StPS. 
\vspace{-1.5mm}

\section{Neural Network Equalizer Design}
\vspace{-1.5mm}

The combination of convolutional and recurrent NNs has proven to work efficiently for speech enhancement~\cite{li2020} and image classification~\cite{choi2017}. In this paper, we analyze the functionality of this advanced NN combination for impairment mitigation in optical transmission systems. 

First, we describe the input layer of our NN equalizers. For the CRNN and biLSTM equalizers, the input shape is a 3D tensor with dimensions $(B, M, 4)$, where $B$ is the mini-batch size, $M$ is the memory size defined as $M=2N + 1$, $N$ being the number of neighboring (past and future) symbols used for the equalization; $4$ is the number of features referring to the real and imaginary parts of two polarization components. For the MLP, the input layer must have a 2D tensor shape: $(B, 4M)$, but the input symbols are the same as for the CRNN and biLSTM.

\begin{figure}[!htb]
    \centering
\begin{minipage}{0.498\textwidth}
\includegraphics[width=\linewidth]{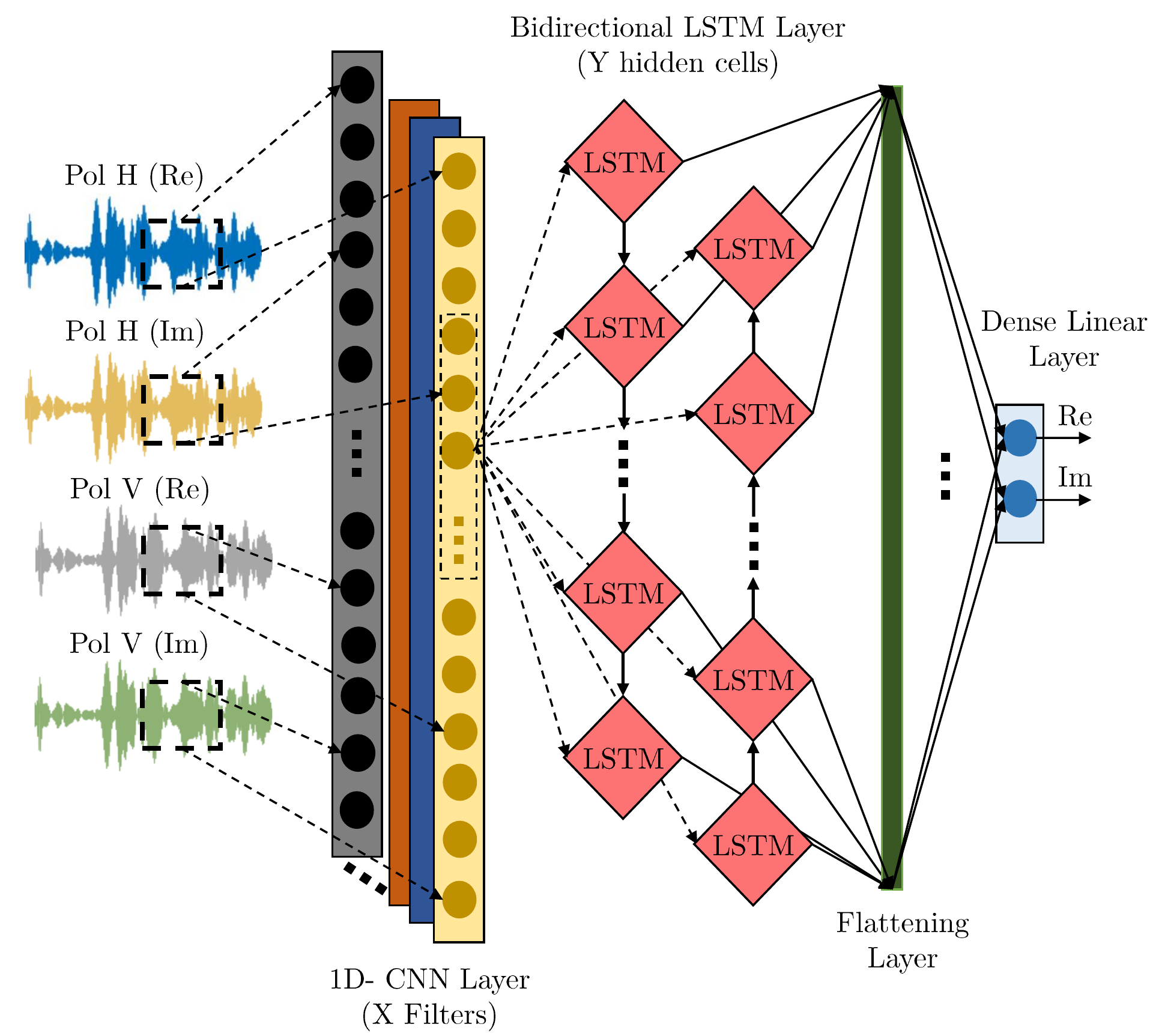}
\vspace{-7mm}
\caption{Scheme of the proposed CRNN equalizer. }\label{fig:figure1}
\end{minipage}
\hfil
\begin{minipage}{0.45\textwidth}
\includegraphics[width=\linewidth]{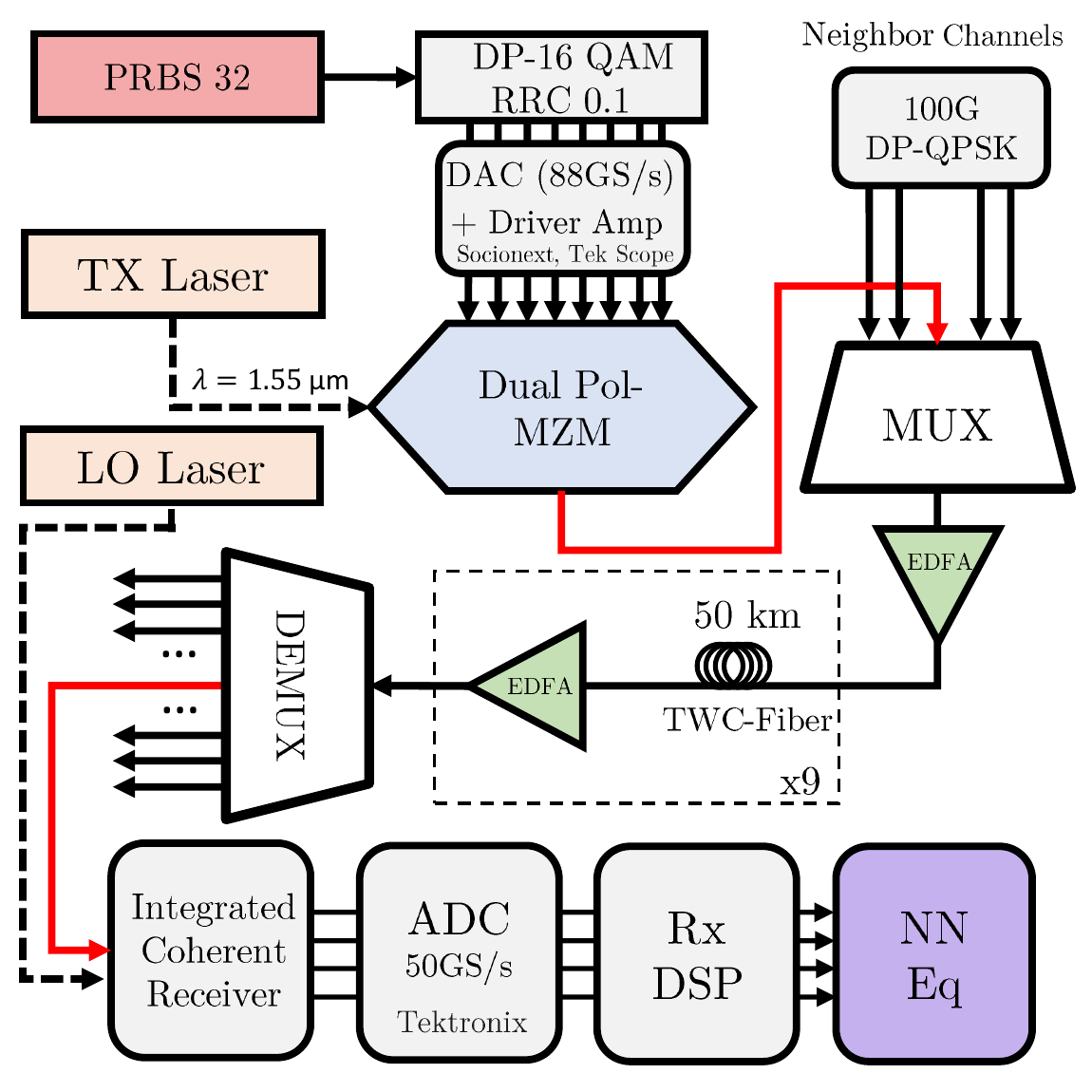}
\vspace{-7mm}
\caption{Experimental setup used for data generation.}\label{fig:figure2}
\end{minipage}
\vspace{-6mm}
\end{figure}

Here we consider three NN topologies: i) two densely-connected layers (i.e., the MLP) equivalent to that used in~\cite{sidelnikov2019methods}; ii) a single biLSTM layer as in~\cite{deligiannidis2020compensation}; and iii) the new CRNN equalizer, with a 1D convolutional layer (1D-Conv) defined by $X$ filters and kernel size $Z$, followed by a biLSTM layer defined by $Y$ hidden units (cells). In the case of CRNN and biLSTM, a flattening layer is used to reduce the output dimensionality. Finally, the output layer with two linear neurons is used in the NN's output to represent the real and imaginary parts of the recovered symbol in one of the polarization. The proposed CRNN equalizer structure is illustrated in Fig.~\ref{fig:figure1}.

Unlike previous approaches, this work implemented a Bayesian optimizer (BO) step, following the procedure described in~\cite{pedro2020}, to define the values of $N$, $X$, $Y$, $Z$, batch size, and the activation function type considered in our CRNN equalizer. 
Having carried out 20 BO cycles, we fix the set of hyper-parameters corresponding to the lowest BER reached. 
The hyper-parameters found by the BO were: $N=20$ taps, $X=244$ filters, the kernel size $Z=3$, $Y=226$ hidden units in the LSTM layer, and the mini-batch size $B=4331$. The activation functions found by the BO for the 1D-Conv and LSTM cells were Leaky ReLU with \(\alpha = 0.2\) and Tanh, respectively. The BO optimization was carried out using the data corresponding to the highest launch power available. For the fair comparison, we also used the same number of hidden units ($Y=226$) found by the BO to the CRNN in the pure biLSTM equalizer and all equalizers used the same number $N$ of taps. For the MLP we set $192$ neurons in the hidden layers since no improvement has been observed when increasing this number further. After having determined the optimal architecture, each NN equalizer was trained for each individual launch power using $2^{20}$ symbols over 200 epochs, employing the traditional MSE loss function and the Adam optimizer with the learning rate of 0.001.  The validation set and evaluation of the resulting BER were carried out using $2^{17}$ independent symbols.
Finally, we highlight that the NN training data were shuffled using numpy.random.shuffle function in Python before feeding it into the NN: such a shuffling eliminates any possible data periodicity. The independent datasets were created using a pseudo-random binary sequence (PRBS) of order 32 to avoid overestimation~\cite{Eriksson2017}. 
\vspace{-1.5mm}

\section{Results and Discussions}
\vspace{-1.5mm}

Fig.~\ref{fig:figure2} shows the setup used in our experiment. At the transmitter, the DP-16QAM 34.4 GBd symbol sequence was mapped out of data bits generated by a $2^{32} - 1$ order PRBS. A digital RRC filter with roll-off 0.1 was applied to limit the channel bandwidth to 37.5 GHz. The filtered digital samples were uploaded to a digital-to-analog converter (DAC) operating at 88 Gsamples/s. The outputs of DAC were amplified by a four-channel electrical amplifier which drove a dual-polarization in-phase/quadrature Mach–Zehnder modulator, modulating the continuous waveform carrier produced by an external cavity laser at $\lambda = 1.55 \mu m$. 
The resulting optical signal was then transmitted along 9$\times$50 km spans of TWC optical fiber with EDFA amplification only together with up to 95 neighboring channels (100G QPSK, 50 GHz ITU grid, for the WDM scenario). The parameters of the TWC fiber span at $\lambda = 1.55 \mu m$  are: attenuation coefficient $ \alpha = 0.23$ dB/km, dispersion coefficient $D = 2.8$ ps/(nm·km), and effective nonlinear coefficient $\gamma$ = 2.5 (W·km)$^{-1}$. 
At the RX side, the optical signal was converted into the electrical domain using an integrated coherent receiver. The resulting  signal was sampled at 50 Gsamples/s by a digital sampling oscilloscope and processed by an offline DSP based on \cite{kuschnerov2010data} which includes chromatic dispersion compensation, MIMO equalization, clock recovery and a pilot-aided carrier recovery. The system performance is evaluated in terms of the Q-factor: $Q = 20 \: \mathrm{log_{10}} \left[\sqrt{2} \: \mathrm{erfc^{-1}}(2\,\rm BER)\right]$.

\begin{figure*}[ht!] 
  \begin{subfigure}[b]{0.5\linewidth}
    \centering
 \begin{tikzpicture}[scale=0.77]
    \begin{axis} [ylabel={BER}, 
        xlabel={Launch power [dBm]},
        ylabel={Q-Factor [dB]},
        grid=both,  
        xmin=-6, xmax=5,
    	xtick={-6, ..., 5},
        legend style={legend pos=south west, legend cell align=left,fill=white, fill opacity=0.6, draw opacity=1,text opacity=1},
    	grid style={dashed}]
        ]
    
    \addplot[color=red, mark=o, very thick]   
    coordinates {
    (-6,7.62)(-4, 8.26)(-3, 8.51)(-2, 8.73)(-1, 8.92)(0, 9.12)(1,9.19)(2,8.85)(3,8.19)(4,7.01)(5,5.20)
    };
    \addlegendentry{\footnotesize{CRNN}};

    \addplot[color=blue, mark=square, very thick]     coordinates {
    (-6,7.60)(-4, 8.28)(-3, 8.48)(-2, 8.598)(-1, 8.73)(0, 8.90)(1,8.70)(2,8.20)(3,7.06)(4,5.70)(5,4.10)
    };
    \addlegendentry{\footnotesize{biLSTM\cite{deligiannidis2020compensation}}};

    \addplot[color=green, mark=triangle, very thick]     coordinates {
        
    (-6,7.59)(-4, 8.24)(-3, 8.46)(-2, 8.56)(-1, 8.56)(0, 8.32)(1,7.93)(2,7.07)(3,6.10)(4,4.47)(5,2.81)
    };
    \addlegendentry{\footnotesize{MLP\cite{sidelnikov2019methods}}};

            \addplot [color=orange, mark=x, very thick]    coordinates {
                    (-6,7.59)(-4, 8.24)(-3, 8.46)(-2, 8.42)(-1, 8.36)(0,8.02)(1,7.71)(2,6.92)(3,5.90)(4,4.15)(5,2.53)
    };
    \addlegendentry{\footnotesize{DBP - 3 StPS \cite{dbp2014}}};

        \addplot[color=magenta,dash pattern=on 1pt off 3pt on 3pt off 3pt,
                mark=*,very thick]    coordinates {
    (-6,7.5)(-4, 8.1)(-3, 8.2)(-2, 8)(-1, 7.8)(0, 7.3)(1,6.5)(2,5.6)(3,4.3)(4,2.4)(5,0.6)
    };
    \addlegendentry{\footnotesize{Regular DSP \cite{kuschnerov2010data}}};

    \end{axis}
    \end{tikzpicture}
    \vspace{-1mm}
    \caption{Single Channel - TWC Fiber (450~km)}
    \label{fig7:b} 
    \vspace{4ex}
  \end{subfigure} 
  \begin{subfigure}[b]{0.5\linewidth}
    \centering
 \begin{tikzpicture}[scale=0.77]
    \begin{axis} [ylabel={Q-Factor [dB]}, 
        xlabel={Launch power [dBm]},
        grid=both,
    	xmin=-6, xmax=2,
    	xtick={-6, ..., 2},
        legend style={legend pos=south west, legend cell align=left,fill=white, fill opacity=0.6, draw opacity=1,text opacity=1},
    	grid style={dashed}]
        ]

    \addplot[color=red, mark=o, very thick]  coordinates {
    (-6,7.72) (-4,8.1 )(-3,8.33 )(-2, 8.37)(-1,7.93)(0,7.33)(1,6.54)(2,5.53)
    };
    \addlegendentry{\footnotesize{CRNN}};

    \addplot[color=blue, mark=square, very thick]     coordinates {
    (-6,7.59) (-4,7.94 )(-3,8.11 )(-2, 7.8)(-1,7.35)(0,6.40)(1,5.20)(2,3.82)
    };
    \addlegendentry{\footnotesize{biLSTM\cite{deligiannidis2020compensation}}};

    \addplot[color=green, mark=triangle, very thick]     coordinates {
    (-6,7.57) (-4,7.70 )(-3,7.50 )(-2, 6.95)(-1,6.10)(0,5.05)(1,3.65)(2,2.12)
    };
    \addlegendentry{\footnotesize{MLP\cite{sidelnikov2019methods}}};
            \addplot[color=orange, mark=x, very thick]      coordinates {
    (-6, 7.40) (-4,7.59 )(-3,7.11)(-2, 6.63)(-1,5.75)(0,4.67)(1,3.32)(2,1.86)
    };
    \addlegendentry{\footnotesize{DBP - 3 StPS \cite{dbp2014}}};
    
        \addplot[color=magenta,dash pattern=on 1pt off 3pt on 3pt off 3pt,
                mark=*,very thick]     coordinates {
    (-6,7.23) (-4,7.33 )(-3,6.86 )(-2, 6.23)(-1,5.33)(0,4.20)(1,2.82)(2,1.22)
    };
    \addlegendentry{\footnotesize{Regular DSP \cite{kuschnerov2010data}}};

    \end{axis}
    \end{tikzpicture}\vspace{-1mm}
    \caption{96 Channels WDM - TWC Fiber (450~km)}
    \label{fig7:a} 
    \vspace{4ex}
  \end{subfigure}
\vspace{-11mm}
  \caption{Performance of the proposed CRNN, benchmarked against biLSTM equalizer~\cite{deligiannidis2020compensation}, two layers MLP equalizer~\cite{sidelnikov2019methods} and DBP 3 StPS \cite{dbp2014} for two experimental fiber-optic scenarios.}
  \label{fig7} \vspace{-5mm}
\end{figure*}
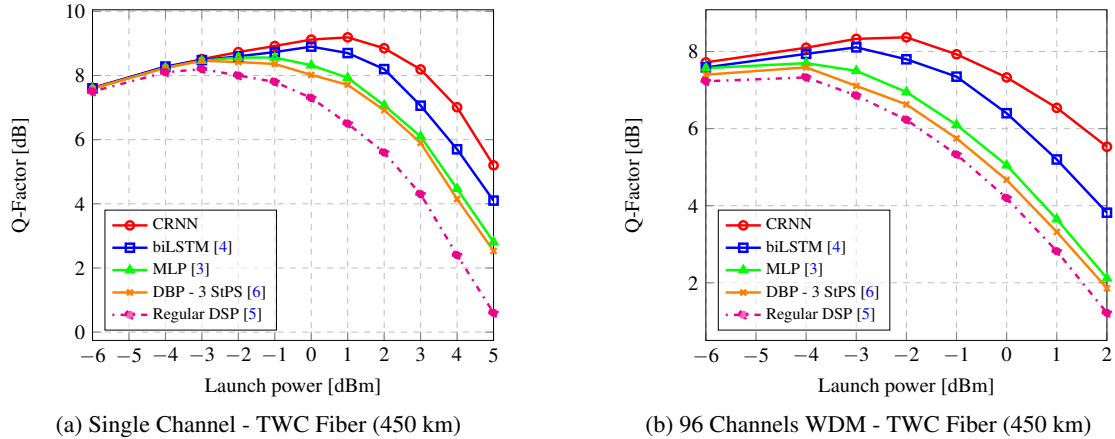

Fig.~\ref{fig7:b} and Fig.~\ref{fig7:a} display the results obtained by the proposed CRNN for the SC and WDM systems, respectively, and their comparison with the results for MLP and biLSTM equalizers. For the SC, the proposed CRNN improved the Q-factor by 1~dB when compared to using the linear DSP only, by 0.29~dB when compared with the biLSTM equalizer, by 0.63~dB when compared to the results of the MLP equalizer, and by 0.73~dB when compared to the DBP with 3 StPS. Additionally, for the CRNN, the optimum launch power improved from -3~dBm to 1~dBm, which highlights the new method's potential to mitigate nonlinear effects. This result may be a consequence of using the convolutional layer before the recurrent layers. By doing so, we extract middle-level locally invariant features from the input series, thus creating a pre-enhancement step for the signal fed into the biLSTM layer. The parameters of the DBP were also optimized to produce the best BER. Importantly, we notice that the result in Fig.~\ref{fig7:b} agrees with the conclusions of previous simulations: the MLP performs slightly better than DBP \cite{sidelnikov2019methods}, and the biLSTM Q-factor improvement is noticeably higher than the DBP one\cite{deligiannidis2020compensation}.  

A 96 channel WDM system was also considered to evaluate the performance of the considered NN equalizers in presence of inter-channel crosstalk. In this case, the CRNN improved the Q-factor by 1~dB when compared to the regular DSP only, by 0.26~dB when compared to the biLSTM equalizer, by 0.67~dB when compared to the  MLP equalizer, and by 0.78~dB when compared to the 3 StPS DBP. The optimum launch power also increased by 2 dB when using the proposed CRNN: from -4~dBm to -2~dBm. Moreover, as can be seen, the topologies that use the biLSTM layers were able to partially recover the XPM impact insofar as some higher gain was observed compared to the results of the SC scenario, Fig.~\ref{fig7:b}. As an example, using the CRNN equalizer at -2 dBm launch power, we observe a Q-factor improvement of $0.73$ dB in the SC case and a $2.14$ dB improvement in the WDM case when compared to the reference Q-factor level (Regular DSP). This effect was also reported by means of numerical simulations in \cite{deligiannidis2020compensation}, where it was stated that LSTM layers are able to ``deterministically'' track sufficiently slow XPM effects in scenarios where no information on neighbor channels is fed into the NN-based equalizer.

\vspace{-1.5mm}
\section{Conclusions}
We proposed a new advanced composite CRNN equalizer to mitigate optical fiber transmission impairments, and compared its performance with several NN-based nonlinear equalizers using experimental data considering both single-channel and  WDM systems.  The new NN topology rendered a significant system performance improvement compared to the results produced by the previously proposed simpler NN architectures, thus demonstrating its potential to mitigate signal distortions arising both from SPM and XPM.
\vspace{-1.5mm}

\section{Acknowledgements}\vspace{-1mm}
\footnotesize
\linespread{0.0}
This work has received funding from: EU Horizon 2020 program under the Marie Skodowska-Curie grant agreement No. 813144 (REAL-NET). JEP is supported by the Leverhulme Project RPG-2018-063. SKT acknowledges the support of the EPSRC project TRANSNET.
\normalsize
\linespread{1.0}

\end{document}